\documentclass[aps, prl,showpacs,amsmath,amssymb,amsfonts,twocolumn,superscriptaddress,floatfix,footinbib]{revtex4-1}
\usepackage{float}
\usepackage{amsmath}
\usepackage{amssymb}
\usepackage[final]{graphicx}
\usepackage{upgreek}
\usepackage[utf8]{inputenc}
\usepackage{lineno}
\usepackage[normalem]{ulem}

\newcommand{\mueV}{\,\upmu\mathrm{eV}}
\newcommand{\mum}{\,\upmu\mathrm{m}}

\newcommand{\distance}{5mm}			
\usepackage{color}

\widowpenalty
\clubpenalty
\usepackage{ulem}

\begin{document}
\title{Carrier density driven lasing dynamics in ZnO nanowires}
\author{M. Wille}
\affiliation{Universität Leipzig, Institut für Experimentelle Physik II, Linnéstraße 5, 04103 Leipzig, Germany}
\author{C. Sturm}
\affiliation{Universität Leipzig, Institut für Experimentelle Physik II, Linnéstraße 5, 04103 Leipzig, Germany}
\author{T. Michalsky}
\affiliation{Universität Leipzig, Institut für Experimentelle Physik II, Linnéstraße 5, 04103 Leipzig, Germany}
\author{R. Röder}
\affiliation{Friedrich-Schiller-Universität Jena, Institut für Festkörperphysik, Max-Wien-Platz 1, 07743 Jena, Germany}
\author{C. Ronning}
\affiliation{Friedrich-Schiller-Universität Jena, Institut für Festkörperphysik, Max-Wien-Platz 1, 07743 Jena, Germany}
\author{R. Schmidt-Grund}
\affiliation{Universität Leipzig, Institut für Experimentelle Physik II, Linnéstraße 5, 04103 Leipzig, Germany}
\author{M. Grundmann}
\affiliation{Universität Leipzig, Institut für Experimentelle Physik II, Linnéstraße 5, 04103 Leipzig, Germany}
\date{\today}

\begin{abstract}
\noindent We report on the temporal lasing dynamics of high quality ZnO nanowires using time-resolved \mbox{micro-photoluminescence} technique. The temperature dependence of the lasing characteristics and of the corresponding decay constants demonstrate the formation of an electron-hole plasma to be the underlying gain mechanism in the considered temperature range from 10\,K to 300\,K. We found that the temperature dependent emission onset-time~($t_{\text{on}}$) strongly depends on the excitation power and becomes smallest in the lasing regime, with values below 5 ps. Furthermore, the observed red shift of the dominating lasing modes in time is qualitatively discussed in terms of the carrier density induced change of the refractive index dispersion after the excitation laser pulse. This theory is supported by extending an existing model for the calculation of the carrier density dependent complex refractive index for different temperatures. This model coincides with the experimental observations and reliably describes the evolution of the refractive index after the excitation laser pulse.
\end{abstract}
\maketitle
\noindent\textbf{{\large 1. Introduction}}\\\\
Micro- and nanostructures are promising building blocks for the fabrication of compact integrated circuits \cite{Huang2005}, e.g.~to enhance the spatial resolution of sensors and for imaging applications \cite{Wu2013, Pan2013}. Furthermore, they can be grown using a bottom up approach, which avoids complicated structuring process steps but guarantees single crystal quality and optimal optical performance~\cite{Huang2001a}. Semiconductor micro- and nanowires are therefore of special interest, since their electrical properties can be modified over a wide range during the growth process \cite{Cui2000,Borgstroem2008, Gutsche2012a}. Furthermore, in the case of an optical application these structures naturally provide all necessities for a laser system: under high excitation the semiconductor material acts as active medium and the Fabry-Pérot type resonator geometry is provided by the end facets of the nanowire. Although the optical properties of zinc oxide~(ZnO) micro- and nanowires have been intensively investigated in the last years \cite{Huang2001,Zimmler2008,Czekalla2008}, the laser dynamics in terms of the temporal dependence of the underlying gain profile, as well as the switch-on characteristics of the nano-emitter itself are not yet investigated in detail over a large temperature range. Recently it has been demonstrated using double-pumped time-correlated photoluminescence (PL) technique, that the switch-on time of semiconductor nanowires is below~5\,ps with lowest values of around $\sim$\,1\,ps for ZnO~\cite{Sidiropoulos2014, Roeder2015}.\\
\noindent\hspace*{\distance}The aim of this work is to study the laser dynamics of highly excited ZnO nanowire emitters in a temperature range from 10\,K to 300\,K using a time-resolved PL technique. The emission onset-time will be studied in a wide temperature range in dependence on the excitation power. Furthermore, we observed in various lasing experiments a spectral red shift of the propagating Fabry-Pérot modes~(FPM) in time after the maximal emission intensity. 
\begin{figure*}[t!]
\includegraphics[width=\textwidth]{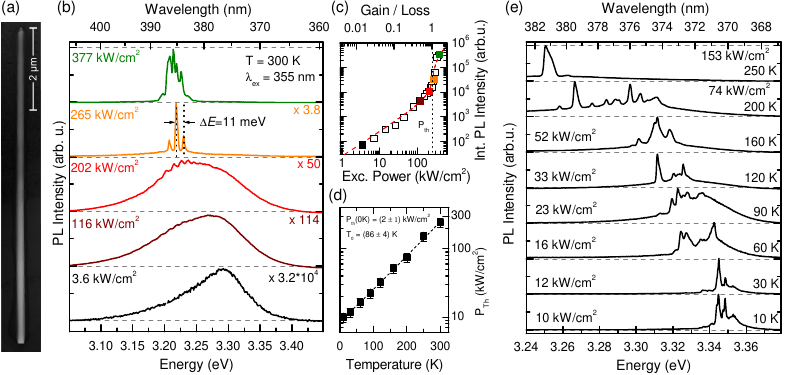}
\caption{(a)~Scanning electron microscope image (SEM) of a ZnO nanowire (length~7.9\,$\mum$, diameter~\mbox{165\,--\,190\,nm}). (b)~The excitation power dependent $\upmu$-PL end facet spectra of the respective nanowire at~300\,K exhibit a transition from broad spontaneous emission to sharp Fabry-Pérot mode~(FPM) emission in the laser regime. (c)~Double-logarithmic plot of the excitation power dependent integrated PL intensity at~300\,K. The power dependence was fitted by a multimodal laser model (red line) with a laser threshold power of $P_{\text{th}}\sim$\,240\,kW/cm$^\text{2}$. The colored squares correspond to the spectra in~(b). (d)~Temperature dependent lasing experiments exhibit an exponential increase of $P_{\text{th}}$ with increasing temperature. (e)~Temperature dependent PL spectra at the respective threshold power exhibit a red shift of the main emission due to the shift of the band gap and a broadening of the laser modes with increasing temperature.}
\label{Lasing}
\end{figure*}
This effect is only accessible due to the Fabry-Pérot type morphology of nanowires and enables the investigation of the time dependent complex refractive index in nanowires. Therefor we extended an existing model for the calculation of the carrier density dependent refractive index and correlated these simulations to our time-resolved measurements.\\\\
\noindent\textbf{{\large 2. Methods and experiment}}\\\\
ZnO nanowires were synthesized by a chemical vapour deposition (CVD) process using the vapour-liquid-solid~(VLS) mechanism \cite{Wagner1964}. A mixture of ZnO- and carbon powder (molar ratio of 1:1) was evaporated in a horizontal tube furnace at 1050\,$^\circ$C and transported by a mixture of Ar and O$_2$ gas towards the silicon substrate, which was beforehand coated with a 200\,nm thick ZnO seed layer. The pressure and growth time were set to  120\,mbar and 30\,min. Single crystalline nanowires grow with diameters in the range of (100\,-\,400)\,nm and lengths up to~50\,$\mum$. Single nanowires were transferred afterwards onto a clean SiO$_2$/Si substrate by dry imprint technique for subsequent $\upmu$-PL measurements. The thermally grown low refractive index SiO$_2$ layer (thickness of 1.5\,$\mum$) ensures the strong optical mode confinement in the nanowire waveguide. Thus, the leakage of the electromagnetic field into the substrate is avoided and hence the energy dissipation out of the nanowire optical cavity~\cite{Roeder2013}. The nanowire sample was mounted in a liquid helium flow cryostat with integrated heating unit enabling investigations in the temperature range from 10\,K to 300\,K. A frequency doubled Ti:Sapphire laser ($\lambda_{\text{ex}}$\,=\,355\,nm, $t_{\text{pulse}}$\,=\,2\,ps, f$_{\text{rep}}$\,=\,76\,MHz) was focused by a 50\,$\times$ NUV microscope objective (NA\,=\,0.4) to a spot size of around 50\,$\mum^2$ for the non-resonant and full-area excitation of single nanowires. A variable attenuator was used to adjust the excitation power density. The luminescence light was collected by the same objective, and than split up into two beams using a non-polarizing beam splitter~(50/50). One beam was dispersed by a spectrometer (320~mm focal length, 2400\,groves/mm grating) and detected by a Peltier-cooled, back-illuminated CCD camera to reach a spectral resolution of~$\sim$\,500$\mueV$. The other beam was dispersed by a spectrometer (320~mm focal length, 600\,groves/mm grating) and detected by a streak camera (Hamamatsu C5680) with a temporal resolution of~$\sim$\,5\,ps. The laser power was measured using a Si diode power meter.\\\\\\
\textbf{{\large 3. Results and Discussion}}\\\\
\textit{3.1. Time-integrated lasing from 10\,K to 300\,K}\\\\
Figure~\ref{Lasing}(a) depicts a scanning electron microscope image of a slightly tapered ZnO nanowire with a diameter ranging from~165\,--\,190\,nm and a length of L\,=\,7.9\,$\mum$. The entire experimental results presented in this work originate from this single ZnO nanowire. Note, that comparable ZnO nanowires exhibit similar optical properties. Figure~\ref{Lasing}(b) shows the emission spectra obtained from a single facet of the ZnO nanowire for several excitation powers at~300\,K. With increasing excitation power the broad spontaneous, excitonic emission out of the entire wire volume transforms into emission spectra dominated by stimulated emission from distinct FPM whose energy is approximated by $E_{Mode} = (h\,c\,N)/(2\,L\,n),$ with $N$ being the modenumber of the FPM, L=7.9\,$\mum$ is the resonator length and $n$ is the carrier density dependent refractive index. At the lasing threshold, the refractive index has a value of $n\sim 2.35$. The precise correlation between the refractive index and the carrier density is discussed below in detail.  Furthermore, the main PL emission exhibits an overall spectral red shift and a mode enhancement on its low energy side with increasing excitation power~\cite{Bohnert1980}. This is caused by the formation of an electron-hole plasma~(EHP) being the underlying gain process in our nano-laser system at~300\,K. The laser threshold power~$P_{\text{th}}$\,$\sim$\,240\,kW/cm$^{\text{2}}$ was obtained by modeling the excitation power dependent integrated PL intensity with a multimodal laser approach \cite{Casperson1975} (red line in Fig.~\ref{Lasing}(c)).
\begin{figure*}[t]
\centering
\includegraphics[width=\textwidth]{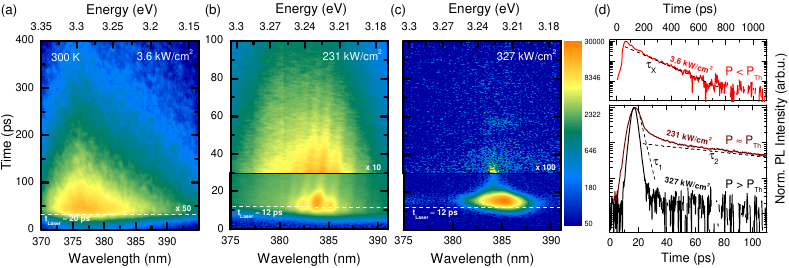}
\caption{Time-resolved PL images (a)~below, (b)~at and (c)~above threshold power~$P_{\text{th}}$ at 300\,K in logarithmic color scale of the single ZnO nanowire shown in Fig.~\ref{Lasing}\,(a). The dashed, white lines indicate the impinging time of the laser pulse. (d)~Normalized transients of the depicted time-resolved PL images taken at the maximum mode intensity. Below~$P_{\text{th}}$ (red line), the lifetime of unscreened excitons~$\tau_{\text{x}}$\,=\,150\,ps is determined by a monoexponential fit. The biexponential model, exemplarily shown at~$P_{\text{th}}$~(brown line), reveals $\tau_1$\,$\lesssim$\,5\,ps (EHP) and $\tau_2$\,=\,(120\,$\pm$\,7)\,ps (screened excitons). Note the different time scale between (a) and (b,c) as well as between the red and the brown and black lines in (d).}
\label{300K_Streak}
\end{figure*}
The evolution of~$P_{\text{th}}$ in the considered temperature range from 10\,K to 300\,K is shown in Fig.~\ref{Lasing}(d). With increasing temperature~$P_{\text{th}}$ increases exponentially according to \mbox{$P_{\text{th}}(T)=P_{\text{th}}(0\,\text{K}) + A \exp(T/T_0)$}, with \mbox{$P_{\text{th}}(0\,\text{K})$\,=\,(2$\,\pm\,$1)\,kW/cm${^\text{2}}$}, \mbox{\textit{A}\,=\,(7.6$\,\pm\,$1.2)\,kW/cm${^\text{2}}$} and a characteristic temperature \mbox{\textit{T}$_0$\,=\,(86$\,\pm\,$4)\,K}. These values are in agreement with the literature~\cite{Fallert2009}. The temperature \textit{T}$_0$ is characteristic for temperature correlated losses in the nano-laser system like the carrier redistribution in k-space due to the change of the Fermi distribution. Thus, high values of \textit{T}$_0$ imply that the threshold power of the device increases less rapidly with increasing temperature and the laser device is thermally more stable.
Figure~\ref{Lasing}(e) depicts temperature dependent PL spectra at corresponding~$P_{\text{th}}$ exhibiting an overall red shift of the band gap energy, a continuous broadening of the underlying gain profile~\cite{Bohnert1980} and a broadening of the FPM with increasing temperature. The large broadening of the dominating FPM at elevated temperatures (see especially the spectra at 250\,K in Fig.~\ref{Lasing}(e)) is correlated to their energetic shift in the first picoseconds after the excitation pulse, as we will show below.\\\\
\textit{3.2. Time-resolved lasing from 10\,K to 300\,K}\\\\
Time-resolved $\upmu$-PL measurements were performed in order to investigate the temporal dynamics of the nano laser system. Figures~\ref{300K_Streak}(a-c) show exemplary time-resolved images of the ZnO nanowire emission in false color for different excitation powers at~300\,K. The broad spontaneous emission (see Fig.~\ref{300K_Streak}(a) and the black curve in Fig.~\ref{Lasing}(b)) exhibits an effective radiative decay time of approximately~$\tau_{\text{x}}$\,=\,(150\,$\pm$\,15)\,ps~(see Fig.~\ref{300K_Streak}(d)). This is a typical value for the average recombination lifetime of unscreened excitons in ZnO~\cite{Jung2002}. Due to the higher generation rate of electron-hole pairs at elevated excitation powers, exciton scattering becomes more favorable. 
\begin{figure}[!b]
\includegraphics[width=0.5\textwidth]{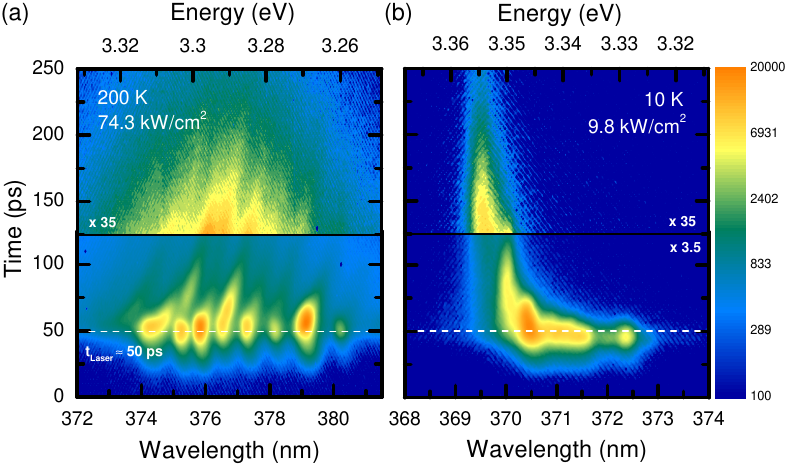}
\caption{Time-resolved PL images at (a)~200\,K and (b)~10\,K at the respective threshold power in logarithmic color scale. The dashed, white lines indicate the impinging time of the laser pulse.}
\label{Tabh_Streak}
\end{figure}
This leads to lower exciton lifetimes and thus to a faster radiative decay time. At further elevated excitation powers another very fast scattering process becomes important, dominating the radiative decay above~$P_{\text{th}}$, see~Fig.~\ref{300K_Streak}(c). 
\begin{figure*}[t]
\centering
\includegraphics[width=\textwidth]{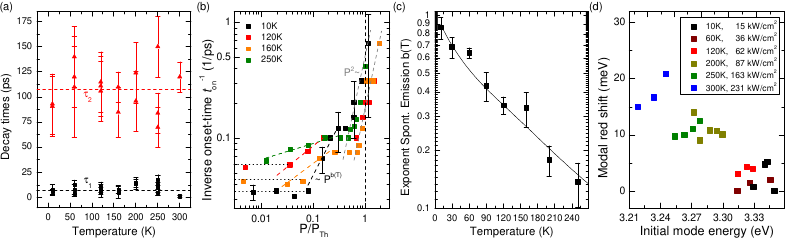}
\caption{(a)~The decay times $\tau_1$ and $\tau_2$ of the biexponential evaluation at P$_{th}$ remain approximately constant in the temperature range from 10\,K to 300\,K. The data points at equal temperature correspond to different FPM. (b)~Inverse emission onset-time~$t_{\text{on}}^{-1}$ of the nano-emitter in dependence of the excitation power for different temperatures. At lowest excitation power the dotted lines mark a regime where $t_{\text{on}}^{-1}$ is independent of the excitation power. In the intermediate regime the increase of the~$t_{\text{on}}^{-1}$ depends strongly on temperature (see dashed colored lines, representing power fits $t_{\text{on}}^{-1} = a P^{b(T)}$). Close to the laser threshold power~$t_{\text{on}}^{-1}$ exhibits a quadratic increase with increasing excitation power (see dashed grey lines). The error bars, which are mainly caused by the temporal resolution of the setup, are only shown for some data points of the 10\,K series for the sake of clarity. (c)~The exponent~b(T) exhibits a strong decreases with increasing temperature. The solid line represents the relaxation model (eq.~\ref{eq:relax}) with fit parameters $\Gamma_0$\,=\,(0.95\,$\pm$0.10)\,meV, $\beta_{\text{ph}}$\,=\,(0.016\,$\pm$\,0.005)\,meV, $\beta_{\text{LO}}$\,=\,(47\,$\pm$\,8)\,meV, E$_{\text{LO}}$\,=\,(72\,$\pm$\,9)\,meV. (d)~Temperature dependence of the absolute modal red shift, extracted in the first 100\,ps after excitation at threshold power.}
\label{Obs}
\end{figure*}
This very fast recombination process originates from the high scattering rates between highly screened electrons and holes in an electron-hole plasma~(EHP). Overcoming the Mott density, the exciton binding energy gets more and more reduced and a fermionic treatment of the available charge carriers is necessary. Figure~\ref{300K_Streak}(d) depicts the transients of the PL decay below (red), at (brown) and above (black) the laser threshold~$P_{\text{th}}$ demonstrating the transition from excitonic emission with a decay constant of~$\tau_{\text{x}}$\,=\,(150\,$\pm$\,15)\,ps to a fast decay triggered by an EHP at high excitation densities ($\tau_1\lesssim$\,5\,ps). The biexponential model of the decaying PL~emission at~$P_{\text{th}}$ demonstrates that the very fast decay ($\tau_1\lesssim$\,5\,ps) changes over to a slower decay of screened excitons with a time constant of $\tau_2$\,=\,(120\,$\pm$\,7)\,ps after the first picoseconds, in which the laser pulse is emitted. This demonstrates the proceeding recombination of carriers and the corresponding relaxation of the residual carrier density in time also after the coherent laser emission. Hence, within some picoseconds the carrier density bears down the Mott density and deceeds it consequently. Thus, subsequently the PL decay is characterized by excitonic recombination, see Fig.~\ref{300K_Streak}(d).\\\\
\textit{3.3. Observations from time-resolved experiments}\\\\
The evaluation of a comprehensive series of time-resolved PL measurements in the temperature range from 10\,K to 300\,K (Fig.~\ref{Tabh_Streak} shows the exemplary time-resolved PL images at 200\,K and 10\,K) led to three main observations.\\
\noindent\hspace*{\distance}\textit{1.~The decay constants} $\tau_{1}\leq$\,5\,ps and $\tau_{2}$\,=\,(75\,--\,150)\,ps do not change significantly as function of temperature, see Fig.~\ref{Obs}(a). Hence, the originating gain mechanism retains over the investigated temperature range. A changing gain mechanism would be accompanied by a significant modification of the main decay time $\tau_1$ as well as a change of the temperature dependence of the threshold power~$P_{\text{th}}$ (Fig.~\ref{Lasing}(d)) would be expected. Both was not observed in our studies. Hence carrier  scattering in an EHP is shown to be the underlying gain process in the ZnO nanowires investigated in this study from~300\,K to~10\,K.\\
\noindent\hspace*{\distance}\textit{2.~The emission onset-time}~$t_{\text{on}}$ is obtained as the time difference between the built-up of the PL intensity to 1/e of its maximum and the temporal maximum of the impinging excitation laser beam. It exhibits a strong dependence on the excitation power as well as on the sample temperature. Figure~\ref{Obs}(b) shows the inverse emission onset-time~$t_{\text{on}}^{-1}$ as a function of the excitation power for different temperatures in a double logarithmic plot. 
\begin{figure*}[t]
\centering
\includegraphics[width=\textwidth]{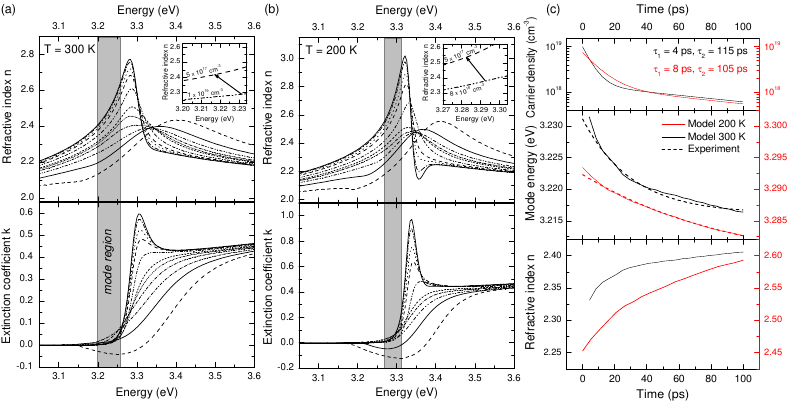}
\caption{Simulation of the refractive index and the extinction coefficient for different carrier densities in ZnO: $1\times 10^{16}\text{\,cm}^{\text{-3}}$~(solid), $1\times 10^{17}\text{\,cm}^{\text{-3}}$~(dash dot), $3\times 10^{17}\text{\,cm}^{\text{-3}}$~(dot), $5\times 10^{17}\text{\,cm}^{\text{-3}}$~(dash), $1\times 10^{18}\text{\,cm}^{\text{-3}}$~(dash dot dot), $2\times 10^{18}\text{\,cm}^{\text{-3}}$~(short dash), $3\times 10^{18}\text{\,cm}^{\text{-3}}$~(short dot), $5\times 10^{18}\text{\,cm}^{\text{-3}}$~(short dash dot), $1\times 10^{19}\text{\,cm}^{\text{-3}}$~(dash dot dot), $2\times 10^{19}\text{\,cm}^{\text{-3}}$~(solid), $4\times 10^{19}\text{\,cm}^{\text{-3}}$~(dash) at~(a)~300\,K and (b)~200\,K. The model parameters in the low carrier density case (10$^{16}$\,cm$^{-3}$) were fitted to temperature dependent ellipsometry data~\cite{Bundesmann2008}. Note the different y-scaling in the diagrams.  The grey boxes mark the energy range where modes appear at and above the respective threshold power. (c)~Time dependence of the carrier density (bi-exponential decay with corresponding decay times $\tau_1$ and $\tau_2$), the calculated mode energy and the respective refractive index at 200\,K (red) and 300\,K (black). The dashed curves in the middle graph represent the experimentally deduced mode shifts. The insets in (a) and (b) depict the initial and final mode energy as well as refractive index, deduced from this calculation.}
\label{Refr_Daempf}
\end{figure*}
By increasing the excitation power, $t_{\text{on}}^{-1}$ increases for all measured temperatures. Thus, the emission sets in faster with increasing excitation power. Far below the laser threshold power ($P\lesssim 0.05\,P_{\text{th}}$) no significant change of $t_\text{on}^{-1}$ is observable, which can be attributed to an inefficient scattering of the excited carriers into the lowest energy states. For a further increase of the excitation power an increase of $t_{on}^{-1}$ can be observed which can be described by a power law ($t_\text{on}^{-1} \propto P^{b(T)}$). The exponent $b$ occupies two excitation regimes; within the intermediate regime ($0.05\,P_{\text{th}}\lesssim P \lesssim 0.5\,P_{\text{th}}$), $b$ is smaller than 1 and depends on the temperature (see Fig.~\ref{Obs}c), following the same temperature dependence as the exciton lifetime,~i.e.
\begin{equation}
b(T)\sim \left(\Gamma_0 + \beta_{\text{ph}}T + \beta_{\text{LO}}/(e^{E_{\text{LO}}/k_BT}-1)\right)^{-1},
\label{eq:relax}
\end{equation}
with $\Gamma_0$, $\beta_{\text{ph}}$, $\beta_{\text{LO}}$ and $E_\text{LO}$ being the inhomogeneous broadening, the exciton-phonon interaction coupling strength, the exciton-LO-phonon interaction strength and the phonon energy \cite{Makino2000}. This relationship between the exponent $b$ and the lifetime indicates a slow relaxation of the excited hot carriers into the ground state so that the carrier relaxation time is larger than the exciton lifetime. Therefore, we can assume that the exciton-phonon scattering is the dominant process even in this intermediate excitation regime. The situation changes for excitation powers $P>0.5\,P_\text{th}$, when the carrier density approaches the Mott density. In this regime, the exponent $b$ changes to a fixed value of $b = 2$, which can be attributed to exciton-exciton and exciton-carrier scattering as the dominant relaxation processes in this excitation range \cite{Arai2006}, accompanied by a strong reduction of the relaxation time. This is also supported by the fact, that the slope of $t_\text{on}^{-1}$ with increasing excitation power in this range is independent of the temperature and therefore independent of the lifetime of the excitons. The dashed vertical line in~Fig.~\ref{Obs}(b) marks the inverse emission onset-time at the respective threshold power being in the range of $t_{\text{on}}^{-1}\sim$\,(0.2\,--0.8)\,ps$^{-1}$. In the laser regime, the corresponding values for~\mbox{$t_{\text{on}}\sim$\,(1.3\,--\,5)\,ps} are limited by the temporal resolution of our setup. Hence, the EHP regime, where carrier-carrier and carrier-phonon scattering are the dominating processes and even shorter onset-times are expected, could not be investigated in this study. For the spontaneous emission, we revealed emission onset-times in the range of~$t_{\text{on}}$\,=\,30\,ps.\\
\noindent\hspace*{\distance}\textit{3.~Mode shift:}
The FPM exhibit a nonlinear spectral red shift in time, whose absolute value increases with increasing temperature from \mbox{1\,meV} at~10\,K up to~\mbox{20\,meV} at~300\,K, see Fig.~\ref{Obs}(d). This modal shift was extracted from the time resolved PL-spectra (e.g. Fig.~\ref{300K_Streak}(b)) in the first 100\,ps after the excitation pulse at threshold power. Each data point in~Fig.~\ref{Obs}(d) represents the red shift of a FPM with a corresponding initial energy evaluated at the time~$t_{\text{on}}$. In general, the mode shift is described by the temporal change of the refractive index dispersion after the optical excitation. Versteegh \textit{et al.} simulated the carrier density dependent dielectric function at room temperature solving the Bethe-Salpeter ladder equation with a matrix inversion method~\cite{Versteegh2011,Haug1990}, including a phenomenological excitonic contribution to the overall carrier density. However, the maximal excitonic fraction of $\sim$\,14\,\% at room temperature leads to a small absolute variation of the resulting refractive index of maximal~$\sim$\,0.3\,\%. Since the treatment of the excitonic contribution in ZnO in that way is questionable and not trivially adaptable to low temperatures, we decided to disregard the excitonic contribution to the overall carrier density in our simulations, being aware that this is a rough approximation calculating the carrier density dependent refractive index. In particular at lower temperatures, the excitonic contribution may play an important role in the intermediate excitation regime. The correlating effects on the refractive index are not considered in this work. Our model parameters were chosen in such a way, that the calculated data reproduce temperature dependent ellipsometry data~\cite{Bundesmann2008} in the low carrier density case ($<$\,10$^{16}$\,cm$^{-3}$). Subsequently, they were held constant for higher carrier densities. Due to a higher value of $k_{max}$, which is necessary for the convergence of the calculated susceptibility, our parameters differ slightly from that in~\cite{Versteegh2011}, with $k_{max}=7\times10^9$\,m$^{-1}$, $d_{cv}=3.2\times 10^{29}$\,C\,m and  $\chi_L$\,=\,0.8. Our simulations display a significant decrease of the Mott density~$n_{\text{Mott}}$ from around $5\times 10^{18}\text{\,cm}^{\text{-3}}$ at~300\,K, $3\times 10^{18}\text{\,cm}^{\text{-3}}$ at~200\,K to around $2\times 10^{17}\text{\,cm}^{\text{-3}}$ at 25~K, which is in good agreement with the model in Ref.~\cite{Klingshirn2006}, based on the description of the electron-hole gas applying Boltzmann statistics. However, due to the disregard of the excitonic contribution to the screening process, the calculated values of~$n_{\text{Mott}}$ has to be treated with caution especially at low temperatures.
Furthermore, because of the uncertainty of the excitonic contribution, we avoided temperatures lower than 200\,K for the calculation of the refractive index. Compared to a nearly negligible excitonic fraction of $\sim$\,14\,\% at 300\,K it becomes already important at 200\,K ($\sim$\,32\%) and tremendous at 100\,K~($\sim$\,87\%).\\
Figure~\ref{Refr_Daempf}(a) and (b) depict exemplarily the simulated refractive index~\textit{n} as well as the extinction coefficient~\textit{k} at 300\,K and 200\,K for different carrier densities. Below the Mott density, the excitonic resonance vanishes with increasing carrier density. The extinction coefficient decreases continuously with increasing carrier density and becomes negative in a certain energy range for carrier concentrations above the Mott density indicating optical gain. The marked grey areas in Fig.~\ref{Refr_Daempf} indicate the energy range, where lasing modes were observed in our experiment. Although the gain profile broadens almost homogeneously, only modes on the low energy side of the gain profile are observed in the experiment. Due to the Gaussian spatial profile of the laser pulse, the nanowire is not homogeneously excited. This may lead to low excited and thus absorbing areas at the nanowire facets causing the absorption of the modes at higher energies. \\
Using this simulations, the experimentally observed mode shift in time after a high optical excitation can be described quite well. The upper graph in Fig.~\ref{Refr_Daempf}(c) depicts the decay of the carrier density in time for lasing modes at 300\,K (black) and 200\,K (red). The corresponding decay times ($\tau_1=4\,\text{ps}$, $\tau_2=115\,\text{ps}$ at 300\,K and $\tau_1=8\,\text{ps}$, $\tau_2=105\,\text{ps}$ at 200\,K) and amplitudes were taken from associated PL transients. We calculated the time dependent refractive index within the nanowires by fulfilling the mode condition. Furthermore we considered, that the nanowire was not excited homogeneously over its entire length because of the Gaussian spatial profile of the laser pulse. For our calculations we chose an excited length of~4\,$\upmu$m. The middle graph in Fig.~\ref{Refr_Daempf}(c) shows that our model (solid lines) describes the experimentally observed temporal mode shift (dashed lines) quite well for 200\,K and 300\,K. Our model exhibits only small deviations in the first picoseconds after excitation. In the considered time range of 100\,ps the calculated refractive index exhibits a continuous increase (see lower graph in Fig.~\ref{Refr_Daempf}(c)), whereas the absolute increase of the refractive index at 200\,K is larger compared to that at 300\,K. This can be explained by considering the mode condition $n E = const$. At low temperatures the strong dispersion causes a strong change of the refractive index when the carrier density decreases. In this case the mode condition can only be fulfilled in a narrow spectral range resulting in a weakly pronounced mode shift. However, at room temperature the broadening of the exciton resonance causes a weaker change of the refractive index but in a wider spectral range. Thus, the experimentally observed stronger mode shift at higher temperatures can be described by our model excellently.\\
Furthermore, we observed at~300\,K that modes closer to the exciton energy exhibit a stronger red shift in time than low energy modes, see~Fig.~\ref{Obs}(d). This is caused by the recovery of the strong spectral dispersion of the refractive index slightly below the exciton energy right after the laser pulse, as can be seen in~Fig.~\ref{Refr_Daempf}(a). Considering the three dominant modes at room temperature (N = 99\,--\,101), we confirmed with our model that the change of refractive index is larger for higher mode energies ($\Delta n_{101} \approx 2.6\,\%$) than for lower mode energies ($\Delta n_{99} \approx 2.2\,\%$, not shown here).
At 200\,K, our model predicts a stronger change of the refractive index as well as an enhanced effect in the proximity of the exciton resonance. Considering three modes around 3.29\,eV, we found $\Delta n \approx 6.6\,-\,7.2\,\%$.
At lower temperatures i.e. 100\,K, the modal red shift is close to the systematic error of the measurement~(0.03\,meV/ps), hence a energy difference for different modes can't be resolved anymore with our streak setup.\\
As mentioned above, the FWHM of the FPM in time-integrated $\upmu$-PL spectra increases at elevated temperatures (see~Fig.~\ref{Lasing}(e)), which is now agreeable explained by the time-resolved dynamics revealing an immense mode shift. At low temperatures the mode shift is negligible, hence the modes appear narrow in the time-integrated spectra. At higher temperatures the mode shift becomes stronger and the modes appear broader. Thus, an evaluation of nanowire Q-values by time integrated spectra is revealed to be rather inaccurate.\\\\\\
\noindent\textbf{{\large 4. Conclusion}}\\\\
In temperature dependent lasing experiments on a single ZnO nanowire, the laser action was found to be stable up to room temperature, whereas the threshold power~$P_{\text{th}}$ exhibits an exponential dependence on the the system temperature with a large characteristic temperature \mbox{\textit{T}$_0$\,=\,(86$\,\pm\,$4)\,K}. Time-resolved PL measurements revealed two characteristic decay constants $\tau_1\lesssim 5$\,ps and \mbox{$\tau_2 = (75-150)$\,ps}, caused by the decay of an electron-hole plasma during the output pulse followed by the excitonic recombination, respectively. A very fast emission onset-time in the lasing regime of~\mbox{$t_{\text{on}}\lesssim 5$\,ps}, below the resolution limit of the used system, was found over the entire temperature range. The experimentally observed temporal red shift of the resonator modes was described by the carrier density driven change of the refractive index in time. Therefor we extended an existing model for the calculation of the carrier density dependent complex refractive index for different temperatures. By merging the decay parameters (decay constants and amplitudes) from time-resolved PL experiments with our model and fulfilling the mode condition in the nano resonator, we calculated the time dependent refractive index after the excitation laser pulse.\\
\noindent\hspace*{\distance}This work was supported by the Deutsche Forschungsgemeinschaft within Gr~1011/26-1, the "Leipzig Graduate School of Natural Sciences~--~BuildMoNa" and through FOR1616. We thank M.\,Ogrisek for the growth of the nanowire samples, as well as M.\,Lorke and M.\,Dorn for valuable discussions.\\\\
\twocolumngrid
\bibliographystyle{apsrev4-1}
\bibliography{Promotion}
\end{document}